# Decoding the Electronic and Structural Fingerprints of Single-Atom Catalysts via DFT-Assisted XANES Analysis

*Petr Lazar,[a] and Michal Otyepka[a,b]\**

[a]Regional Centre of Advanced Technologies and Materials, Czech Advanced Technology and Research Institute (CATRIN), Palacký University Olomouc, Šlechtitelů 27, 783 71 Olomouc, Czech Republic

[b]IT4Innovations, VŠB – Technical University of Ostrava, 17. listopadu 2172/15, Ostrava-Poruba 708 00, Czech Republic

\* Corresponding author: Michal.Otyepka@upol.cz

KEYWORDS: Copper; XANES; single atom catalysis; structure; characterization; graphene

ABSTRACT: Single-atom catalysts (SACs), composed of isolated metal atoms dispersed on solid supports, represent the ultimate expression of atomic efficiency in catalysis. Their remarkable activity and selectivity arise from local coordination environments and adjustable oxidation states, yet precise determination of these features remains an enduring challenge. Among modern characterization techniques, X-ray absorption near-edge structure (XANES) spectroscopy stands out for its sensitivity to both electronic and geometric structure, though its interpretation is often constrained by empirical comparison with bulk references. Here we introduce a density functional theory–based computational spectroscopy framework for the quantitative interpretation of Cu K-edge XANES spectra. The approach accurately reproduces experimental edge positions and fine-structure features across benchmark copper systems, including metallic Cu, $Cu_2O$, CuO, and $CuSO_4 \cdot 5H_2O$. We then employ this framework to reveal the oxidation state, coordination geometry, and hydration environment of Cu single atoms supported on cyanographene, demonstrating direct correspondence between spectral signatures and atomic-scale structure. This methodology establishes a robust and transferable route for connecting XANES features with the underlying electronic and structural characteristics of SACs, thereby advancing the rational design of atomically precise catalysts.

INTRODUCTION

Single-atom catalysts (SACs) represent a cutting-edge class of heterogeneous catalysts, defined by isolated metal atoms dispersed and stabilized on a solid support.[2, 3] These atomically dispersed catalytic centers often outperform their bulk or nanoparticulate counterparts due to their maximized atomic efficiency and unique electronic properties.[4] In SACs, every metal atom acts as an active site, embodying the principle of single-atom economy. The nature of the support material plays a crucial role in stabilizing these atoms in specific oxidation and coordination states, which in turn significantly influence catalytic performance. However, not all these states are catalytically active, making it essential to accurately characterize the structural and electronic environment of the active sites for rational catalyst design.[5-8]

Experimental identification of atomically dispersed species, their oxidation states and chemical environment has remained a major challenge in SAC characterization.[9] Advanced techniques, particularly spectroscopic and microscopic methods, are required to probe SACs at the atomic level. Among them, X-ray absorption near-edge structure (XANES), also referred to as near-edge X-ray absorption fine structure (NEXAFS), has become a cornerstone for confirming the single-atom nature of catalysts and differentiating them from bulk or nanoparticulate forms.[9-11] When complemented by X-ray photoelectron spectroscopy (XPS), transmission electron microscopy (TEM), and scanning transmission electron microscopy (STEM), XANES provides critical insights into oxidation states, coordination environments, and morphological features of catalytic sites. Typically, XANES spectra are interpreted through comparison with known reference compounds. However, transferring this knowledge to novel systems is not straightforward as this approach is challenging when the composition and structure of the material cannot be readily identified. The relationship between structural and chemical features and their spectroscopic signatures remains poorly and empirically understood.

Copper containing SACs have been used for catalysis of $CO_2$ to multi-carbon hydrocarbons with relatively high Faradaic efficiency,[12,13] for hydrogen evolution with high rate and quantum efficiency,[14] and for the oxidative coupling of amines and the oxidation of benzylic C-H bonds toward pharmaceutical applications.[15] Very recently, copper based SACs have provided efficient, selective, and recyclable catalysis of C–S cross-coupling reactions which is essential for producing thioethers used in pharmaceuticals, and agrochemicals.[16] Thus, copper based SACs are emerging in ever growing number of reactions and applications, making it essential to clearly understand the role of Cu single atoms in these processes. Copper can exist in three major oxidation states (Cu(II), Cu(I), or Cu(0)), each potentially contributing differently to catalytic activity. Moreover, the oxidation state, reactivity, and selectivity of Cu-based catalysts can be tuned by the distribution of binding sites for copper. These considerations highlight the importance of developing a deeper understanding of copper K-edge XANES spectra in SACs, particularly signatures associated with different Cu oxidation states and effect of various binding environments.

*Ab initio* computational spectroscopy offers powerful support for interpreting experimental data. Density functional theory (DFT) allows calculation of various spectroscopic properties from first principles. DFT simulations have already proven invaluable for interpreting XPS and vibrational spectra in two-dimensional materials, such as in identifying nitrogen bonding environments in N-doped graphenes,[17] C-F bonding in fluorinated graphenes,[18] or oxygen- and hydrogen-based functionalities in graphene.[19] Also for single-atom systems, DFT-assisted IR spectroscopy has become an effective tool to identify active sites, positions of single atoms on metal oxide surfaces, and their structural evolution in experimental conditions.[10, 20-23] First principles calculations have been also used to identify XANES spectral features of bulk phase, however, analogical studies focused on features of SACs have not been caried out yet. Therefore, reliable indicators that can be used to distinguish properties such oxidation state and coordination shells of single atoms of SACs are missing. Owing to that, the structural and electronic features of single atoms are deduced empirically from fitting measured spectra of SACs to *ad hoc* chosen reference systems naturally having vastly different chemistry.

To fill the gap and establish basic workflow of XANES computational spectroscopy of SACs, we employ quantum mechanical DFT calculations to identify spectroscopic signatures of copper single atoms in graphene-supported SACs. We develop and validate a theoretical framework based on DFT calculations to establish relationships between the structural features of Cu-based single-atom systems and their manifestations in XANES spectra. As a first step, we test the reliability of this framework by simulating XANES spectra for a set of standard Cu reference compounds, including Cu foil (Cu(0)), $Cu_2O$ (Cu(I)), CuO (Cu(II)), and solid-state $CuSO_4 \cdot 5H_2O$. The copper sulphate serves as a representative model for single-atomically dispersed $Cu^{2+}$ ions, which reside in hexa-aqua coordination sphere and lack close Cu-Cu contacts as those existing in bulk Cu and both Cu oxides. We evaluate also influence of additional Hubbard U term which is considered for improved description of strong correlations among Cu *d* electrons in CuO. Our results demonstrate that properly set-up calculations reproduce excellently available experimental spectra, including absolute position of K-edge on the energy axis and various spectral features up to 40 eV from the edge. We then apply the protocol to investigate Cu species anchored on cyanographene.[15] Specifically, we evaluate whether XANES can resolve (i) the oxidation state of Cu, (ii) coordination geometry of Cu bound to the cyanographene surface, (iii) and coordination sphere of Cu, particularly formation of aqua complexes. Our results provide a comprehensive assessment of the capabilities and limitations of XANES in characterizing single-atom catalysts based on copper supported by graphene derivative. We offer computational protocol, which can be used to relate XANES spectral features with structural and electronic properties of Cu-based SACs. The developed workflow can also be translated to other SACs and enables a new method for interpretation of XANES spectra of SACs making important step toward robust characterization of SACs and, hence, understanding of SACs properties, which is required for future rational design of this class of catalysts.

METHODS

All calculations were performed using the projector-augmented wave method as implemented in the Vienna ab-initio simulation package (VASP).[24, 25] The potentials used in the present work are the latest norm-conserving GW potentials including semi-core states (ver. 57). The energy cutoff for the plane-wave expansion was set to 500 eV. We used dense *k*-point grids corresponding to an equivalent of at least 12×12×12 mesh for the primitive cell. The exchange-correlation functional was approximated by the optimized van der Waals DFT functional (optB86b-vdW) which seamlessly includes non-local correlation effects.[26] Additionally, we included a Hubbard term U (U= 4 eV) to the optB86b-vdW functional to improve description of strongly correlated systems such as CuO (see Results and discussion for more details). XANES spectra were calculated using the supercell core-hole method: the Cu 1s core electron was promoted into the conduction bands and matrix elements between core and conduction states were calculated to obtain the absorption spectrum of the core electron. The periodic supercells were used to minimize the spurious interaction between core-holes in neighboring cells. In addition, the number of unoccupied states was increased significantly to obtain the full spectrum within the energy range of up to 40 eV from the absorption edge. This number depended on the supercell size and volume of vacuum, in general around 4000 empty states (bands) above the Fermi level were needed for the supercells used. Notice that the computation time increases drastically with the number of bands, so the computational costs are much higher than those of standard ground-state DFT. It should be mentioned that other approaches to XANES theoretical calculation exist such as real-space multiple-scattering codes FDMNES and FEFF. FEFF uses a real-space Green's function formalism within multiple-scattering theory and is efficient for disordered or cluster systems due to lack of periodic boundary conditions.[27] FDMNES can operate either within multiple-scattering theory or using a finite-difference method to solve the Schrödinger equation, and it allows flexible treatment of core-hole effects and non-spherical potentials.[28] In comparison, the supercell approach within VASP can be computationally more demanding for large cells but provides a fully self-consistent DFT description of XANES within the periodic boundary conditions, allowing consistent treatment of extended systems alongside local cluster model. Recently, these codes have been benchmarked on Ni K-edge XANES of NiO and LiNiO$_2$, and VASP together with FEFF had the highest similarity with the experimental spectra.[29] Given that VASP can be used also for structural relaxations, we adopted it for all subsequent XANES computations.

The XANES spectrum was extracted from the imaginary part of the frequency-dependent dielectric function. The presented spectra were computed as the sum of all components of the dielectric matrix. In core-hole calculation of the dielectric function, a low value of Lorentzian broadening (CH_SIGMA in VASP) of 0.001 eV was used. Then, the obtained XAS spectra were convoluted with a Gaussian function with a full width at half maximum of 1.0 eV to simulate instrument broadening and with a Lorentzian function with a width linearly increasing from 1.0 to 4.0 eV to account for the core-hole lifetime broadening. It should be stated that the observed broadening is driven by many factors and depends on the particular experimental setup, so it is not possible to reproduce the broadening exactly. Calculated discrete spectra were broadened by a Lorentzian function having the width of 1.5 eV to approximate spectral broadening in experiment. Post-processing and broadening of the spectra we done using python *xas-tools* package.[30]

The absolute values of the experimental peak positions are not captured due to limitations of PAW potentials to describe the core level energies. Common practice is either to plot the spectrum on the relative scale starting from zero energy or to align theoretical spectrum to match the onset of experimental Cu K-edge spectrum. However, pre-edge peaks and shoulders may be falsely interpreted when their exact position is not known, hampering meaningful comparison between chemically or structurally different systems such as single copper atoms and copper oxides. Therefore, we adopted an elaborate protocol suggested by England and coworkers[31] in which the absolute position at the energy axis is set through alignment for reference spectrum once and then consistently transferred to other systems.

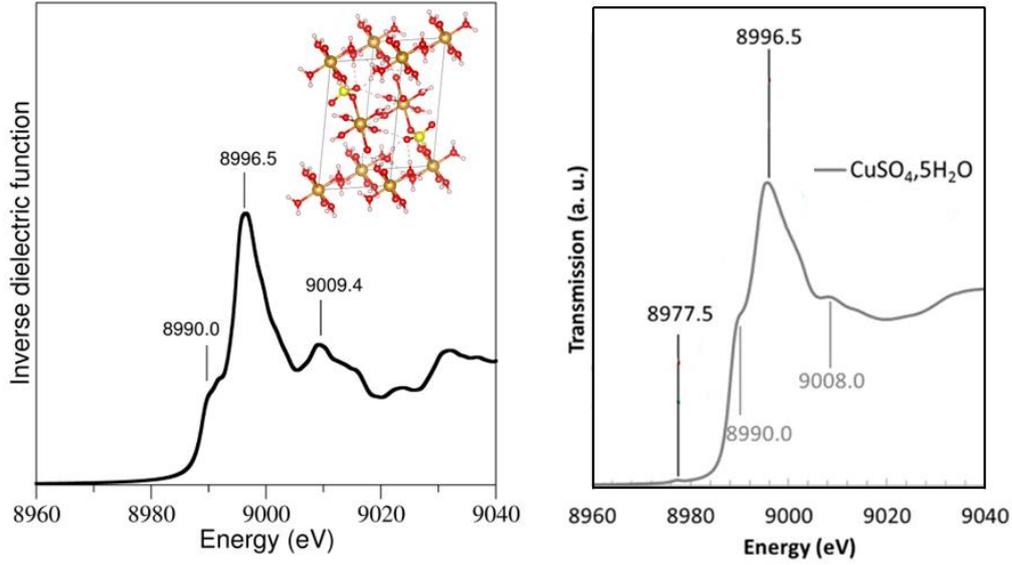

**Figure 1**: Simulated XANES of solid-state $CuSO_4·5H_2O$ (left panel) compared to the experimental XANES spectrum of solid state $CuSO_4·5H_2O$ adopted from Ref.[32] (right panel). The inset shows relaxed structure of crystalline $CuSO_4·5H_2O$ used to calculate XANES spectrum.

Since the protocol by England and coworkers[31] was established to analyze XANES of $CO_2$ molecule, we modified it slightly to fit periodic models of solids. For a given system (Cu SAC, Cu oxide or any other compound), we shift the energy scale as follows:

$$E \rightarrow E - E_F^{CH} + \Delta E_{\text{atomic}} + \Delta_{\text{expt}},$$

where $E_F^{CH}$ is the Fermi energy of the system with core-hole, corresponding to the Kohn–Sham eigenvalue of the first available state that the excited electron can occupy, $\Delta E_{\text{atomic}}$ is the relative excitation energy of given material with respect to that of an isolated Cu atom (we only consider copper 1s core-level excitations in this paper) which is calculated as

$$\Delta E_{\text{atomic}} = \left[E_{\text{system}}^{CH} - E_{\text{atom}}^{CH}\right] - \left[E_{\text{system}}^{GS} - E_{\text{atom}}^{GS}\right],$$

where in parentheses there are differences in the total energy between the system and the isolated Cu atom in ground and excited states. The alignment with experiment is performed once and only once for a reference calculation through the constant $\Delta_{\text{expt}}$ and all other simulated spectra are aligned using the same value of this constant. In our case, we use the XANES of solid-state $CuSO_4 \cdot 5H_2O$ as the reference calculation, because its calculated spectrum has almost one-to-one correspondence with experimental spectrum reported in the literature[32] (Figure 1), which makes the alignment free from any ambiguities. The spectrum dominated by a single peak having maximum at 8996.5 eV, accompanied by low-energy shoulder on the dominant peak and a weaker structure at ~9010 eV. All spectral features are clearly reproduced in simulated spectrum, thus the constant $\Delta_{expt}$ is obtained by shifting calculated spectrum so that the maxima of the dominant peak coincide (are at 8996.5 eV). The obtained value of $\Delta_{expt}$ is then applied to shift all other Cu K-edge XANES throughout the manuscript.

RESULTS AND DISCUSSION

*Simulated XANES Spectra of Reference Systems*

For the Cu(0) oxidation state, Cu foil is the standard and widely applied reference system. Raw experimental data for XANES spectrum of Cu metal foil at 10K are publicly available at X-ray Absorption Data Library, so it represents perfect test system for gauging accuracy of our computational approach. Figure 2a demonstrates that the spectral features and shapes calculated by both optB86b-vdW and optB86b-vdW+U functional are in excellent agreement with experimental data (in this case only, we aligned the calculated spectra to the experimental absorption onset to facilitate the comparison). The simulated XANES spectra reproduce all characteristic features in the copper K-edge spectrum. The near edge part of the absorption spectrum of the bulk Cu exhibits four main features (Figure 2c); A, the transition from 1s to 4p in the $3d^{10}$ configuration, two characteristic peaks of similar intensity, B and C. The fourth peak D at ~ 9026 eV is composed of two peaks in proximity, which are smeared into single broad peak due to the scattering effects. All of these features arise from 1s-4p dipole-allowed transitions. The addition of the Hubbard term U to the functional is not detrimental to the spectrum of metallic copper, on contrary, the term even slightly improves the shape of the spectrum when comparing to the experimental reference (Figure 2a). We therefore applied the optB86b-vdW+U functional throughout all following calculations to keep them on the equal footing and correctly capture the magnetism and electronic structure of CuO (see below).

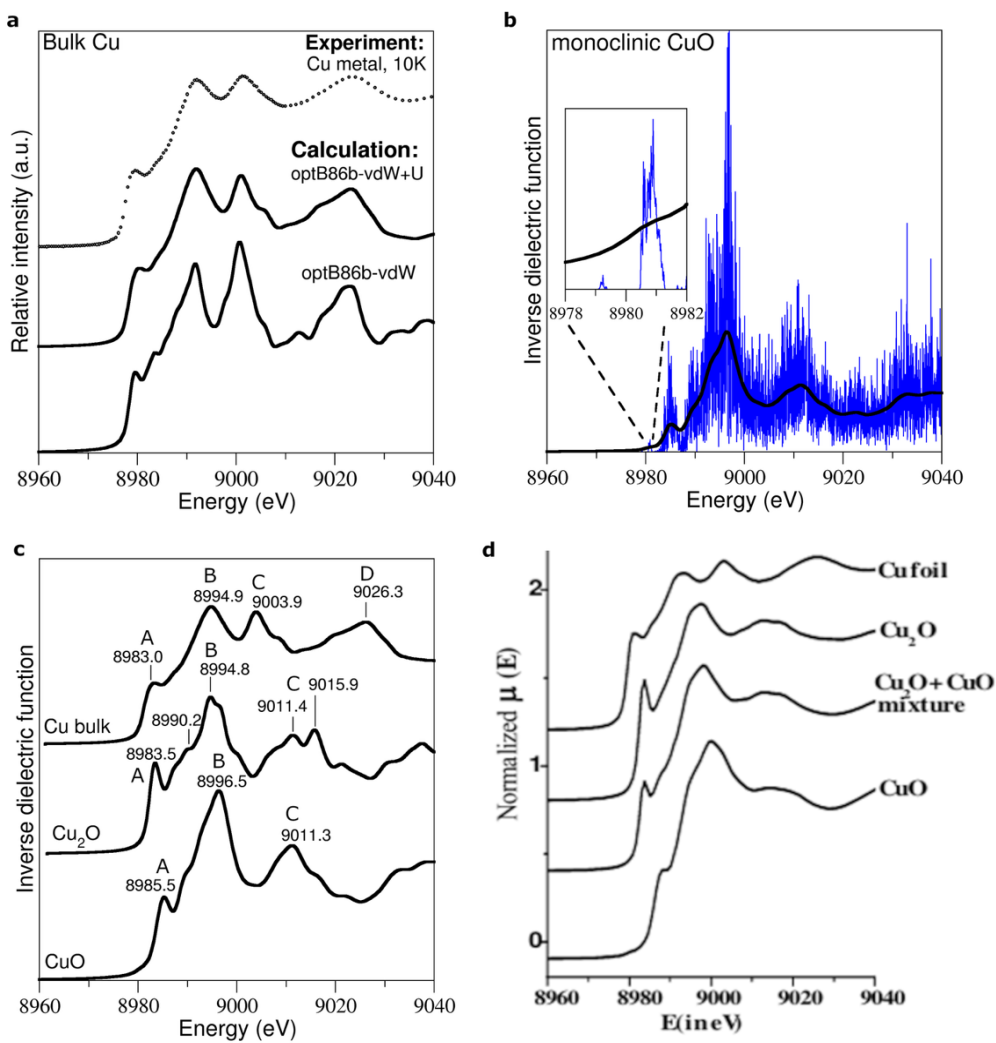

**Figure 2**: XANES spectra of standard Cu references, **a** XANES spectrum calculated w/o the Hubbard U term compared to experimental raw XANES data for Cu metal foil at 10K (taken from X-ray Absorption Data Library), **b** XANES of monoclinic CuO showing raw calculated data (blue) and simulated signal after broadening (black curve), **c** theoretical spectra for bcc Cu, monoclinic CuO, and cubic cuprite $Cu_2O$ compared to, **d**) the experimental spectra reproduced from Ref.[1]

CuO represents one of prototypical solid-state reference models, where Cu atoms are supposed to hold an oxidation state II. CuO crystallizes in a monoclinic face-centered structure (space group: C2/c) with four atoms per unit cell, and the closest Cu-Cu distance is 2.90 Å. CuO is a semiconducting material with an indirect band gap of about 1.2 eV.[33] Below $T_N$ = 213 K, CuO is an antiferromagnet with magnetic moments of ~0.65 $\mu_B$ at the Cu atoms.[34, 35] The correlated nature of CuO presents a great challenge for electronic structure calculations, and standard DFT calculations incorrectly yield CuO as a nonmagnetic metal.[36] The opening of a band gap and the correct anti-ferromagnetic ground state is obtained in LDA+U[37] and with hybrid functionals,[38] but there are many aspects of electronic features of CuO that have not yet been fully settled.[39] These findings and the need to calculate XANES for the physically correct phase of CuO (antiferromagnetic Mott/charge transfer semiconductor) were motivation for an inclusion of the Hubbard term U to the optB86b-vdW functional. By setting U=4 eV, the resulting functional performs very well delivering calculated lattice parameters of CuO *a*=4.65 Å, *b*=3.38 Å, and *c*=5.11 Å matching with experiment[40] (*a*=4.65 Å, *b*=3.38 Å, and *c*=5.11 Å). It reproduces also the antiferromagnetic ordering with local magnetic moments of 0.61 $\mu_B$ at copper atoms and yields CuO as semiconductor with the band gap of 1.34 eV.

The XANES spectrum calculated for antiferromagnetic CuO shows three clear features; A: weak low-energy peak at 8985.5 eV appearing as a shoulder in experimental spectra (Figure 2d), B: strong peak at 8996.5 eV, and c: high energy peak at 9011.3 eV (Figure 2c). The origin of the A peak (shoulder) is quite controversial,[41] because some authors claim that it originates from multiple scattering effects,[42] other ascribe it many-body shake-down processes.[43] In addition to these obvious features, a broad low energy tail below 8985 eV hides a weak pre-edge peak which can be, however, seen in the raw calculated data (Figure 2b). This peak arises from dipole-forbidden but quadrupole-allowed transitions from Cu 1s to empty part of 3d states.[43] This transition occurs only when copper is in the Cu(II) oxidation state, so this weak pre-edge feature can serve as significant fingerprint.

$Cu_2O$ as a prototypical Cu(I) compound exhibits A: distinct low energy peak at 8983.5 eV, B: prominent peak at around 8995 eV and C: double maximum at high energy. Although the overall shape of the spectrum is similar to that of CuO, there appear several notable differences: the edge onset is at lower energy (8983.5 eV vs. 8985.5 eV for CuO), the first peak/shoulder is more pronounced, and the maximum C at ~9013 eV has clear double maximum character, which is resolved also in the experimental spectrum (Figure 2d). It remains to be found to which extent are these differences transferable to single-atom Cu and whether they can be used to deduce the oxidation state of Cu in SAC form. We elaborate this topic further in the following section.

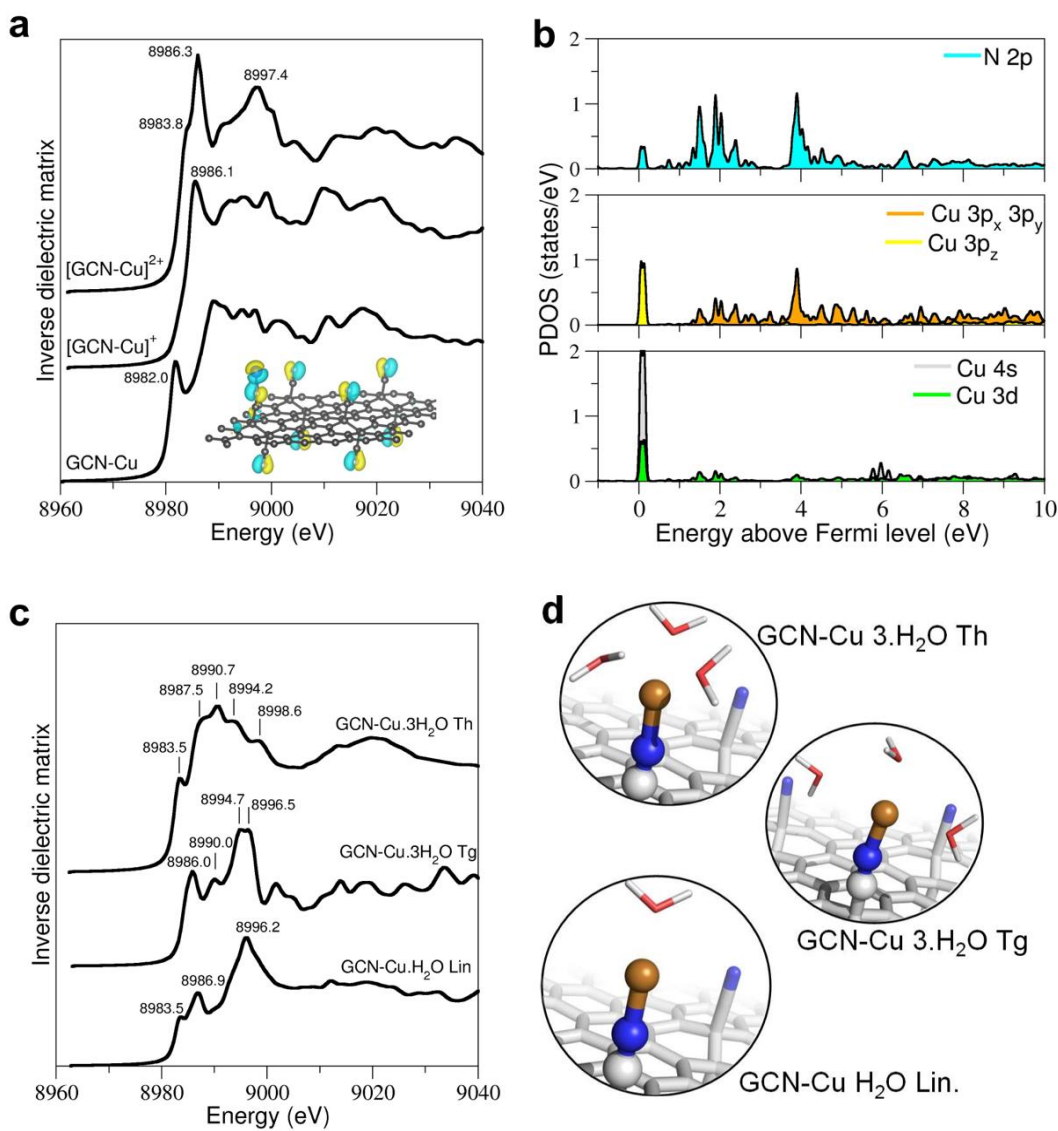

**Figure 3:** Simulated XANES spectra of GCN-Cu SAC. **a**: XANES spectra of GCN-Cu for various charge states (neutral, +e, +2e) imposed on the system, inset**:** the charge density difference between [GCN-Cu]$^{2+}$ and neutral [GCN-Cu]$^{+}$, b: projected density of states for [GCN-Cu]$^{+}$, **c**: XANES spectra of GCN supported Cu surrounded by water molecules in various coordinations, **d:** structures used for XANES simulation of hydrated GCN-Cu.

*Cu single atoms on cyanographene*

In the paper reporting synthesis of GCN-Cu, GCN was mixed with $CuCl_2$ to immobilize single copper ions via coordination to nitrile groups grafted on graphene surface.[15] Although $Cu^{2+}$ ions were immobilized, XPS analysis revealed that roughly half of them were reduced to Cu(I) oxidation state due to charge-transfer from GCN support. This partial reduction was further confirmed by XANES measurements. These findings provide strong motivation for theoretical calculations to better understand the observed reduction processes through simulated XANES spectroscopy.

As a model system of GCN-Cu, we consider a single Cu atom coordinated to one of the nitrile groups on the GCN surface. The calculated XANES spectrum of GCN-Cu is characterized by a dominant peak at ~8986 eV and weak features at higher energies (Figure 3a). The projected density of states (PDOS) reveals that this dominant peak arises from the transition to Cu $p_z$ states, which are however hybridized with the Cu 4s state and partly also with the $p_z$ orbital of anchoring nitrogen (Figure 3b). The calculated spectra indicate that both [GCN-Cu]$^+$ and [GCN-Cu]$^{2+}$ exhibit nearly identical XANES features (Figure 3a). We use square brackets to emphasize that the charge state of an individual anchored ion cannot be controlled directly - only the total charge of the entire system can be adjusted, in accordance with principles of quantum mechanics. The charge density difference between [GCN-Cu]$^{2+}$ and [GCN-Cu]$^+$ shows that the additional charge is delocalized from the Cu atom over neighboring nitrile groups on GCN (inset of Figure 3a). As a result, the formal oxidation state of Cu remains close to Cu(II), regardless of whether the total charge of the system is 1+ or 2+; the Hirshfeld charges on Cu are 0.32 *e*, 0.57 *e*, and 0.67 *e* for the neutral GCN-Cu, [GCN-Cu]$^+$, and [GCN-Cu]$^{2+}$, respectively. For comparison, Cu(II) in solid-state $CuSO_4 \cdot 5H_2O$ has a Hirshfeld charge of 0.59 *e*. Notice that even for the neutral system, the anchored Cu atom is close to the oxidation state I rather than being zerovalent. This interpretation is corroborated by the PDOS analysis, which reveals that most of the Cu 4s electron density lies above the Fermi level – that is, within the unoccupied states.

It is important to note that the spectra shown in Figure 3a were calculated for gas-phase GCN-Cu and neglecting coordination of other ligands, e.g., water. Previous studies reported a significant influence of solvation effects on the characteristics of GCN-metal ion systems. Zaoralová *et al.*[44] demonstrated that an aqueous environment substantially decreased the binding affinities of metal atoms and cations anchored on GCN. Průcha and coworkers[45] showed that variations in the coordination number and spatial arrangement of ligands can alter the oxidation state of a metal atom. To explore these effects, we investigate role of water coordination on the oxidation state and corresponding XANES response of the anchored Cu atom. Three representative configurations are considered based on the most stable geometries reported by Průcha and coworkers[45]; tetragonal (Tg) and tetrahedral (Th) coordination by three water molecules, identified as the most favorable for $Cu^{2+}$, and a linear $[Cu.H_2O)]^+$ complex (Lin), preferred for the $Cu^+$ configuration (Figure 3d). The calculated spectra reveal that water coordination has a profound effect on the XANES features (Figure 3c). The Th complex exhibits a low energy pre-edge peak at 8983.5 eV followed by dominant absorption composed of several peaks centralized around 8991 eV (Figure 3c). The spectrum of Tg complex differs both in the peak shape and position – the onset is shifted to 8986 eV and dominant absorption comes in the form of narrow double peak at 8994.7 and 8996.5 eV. The spectrum of the Lin complex exhibits low-energy peak at 8986.9 eV with the shoulder at 8983.5 eV and single dominant peak at 8996.2 eV. It should be noted that the calculated spectrum of the Tg complex shows the best agreement with the experimental XANES data reported by Bakandritsos and coworkers.[15] The corresponding Hirshfeld charge on the Cu atom in $[GCN-Cu.3H_2O]^{2+}$ (Tg) is 0.43 *e*, indicating reduction compared to the uncoordinated Cu in $[GCN-Cu]^{2+}$ (0.67 *e*), which seems to offer plausible explanation for reduction of $Cu^{+2}$ ions on GCN observed experimentally.[15] It should be noted that, since aqueous solvents are liquids, phenomena such as thermal fluctuations in coordination geometry, structural disorder, and interactions with the extended (band) states of the solvent can influence the spectra measured for real experimental systems.[46] In this context, La Penna and coworkers demonstrated using the archetypal case of $Cu^{2+}$ in water that these effects can be incorporated by evaluating the theoretical spectrum as an average over the computed spectra from a statistically representative ensemble of simulated metal-site configurations.[47]

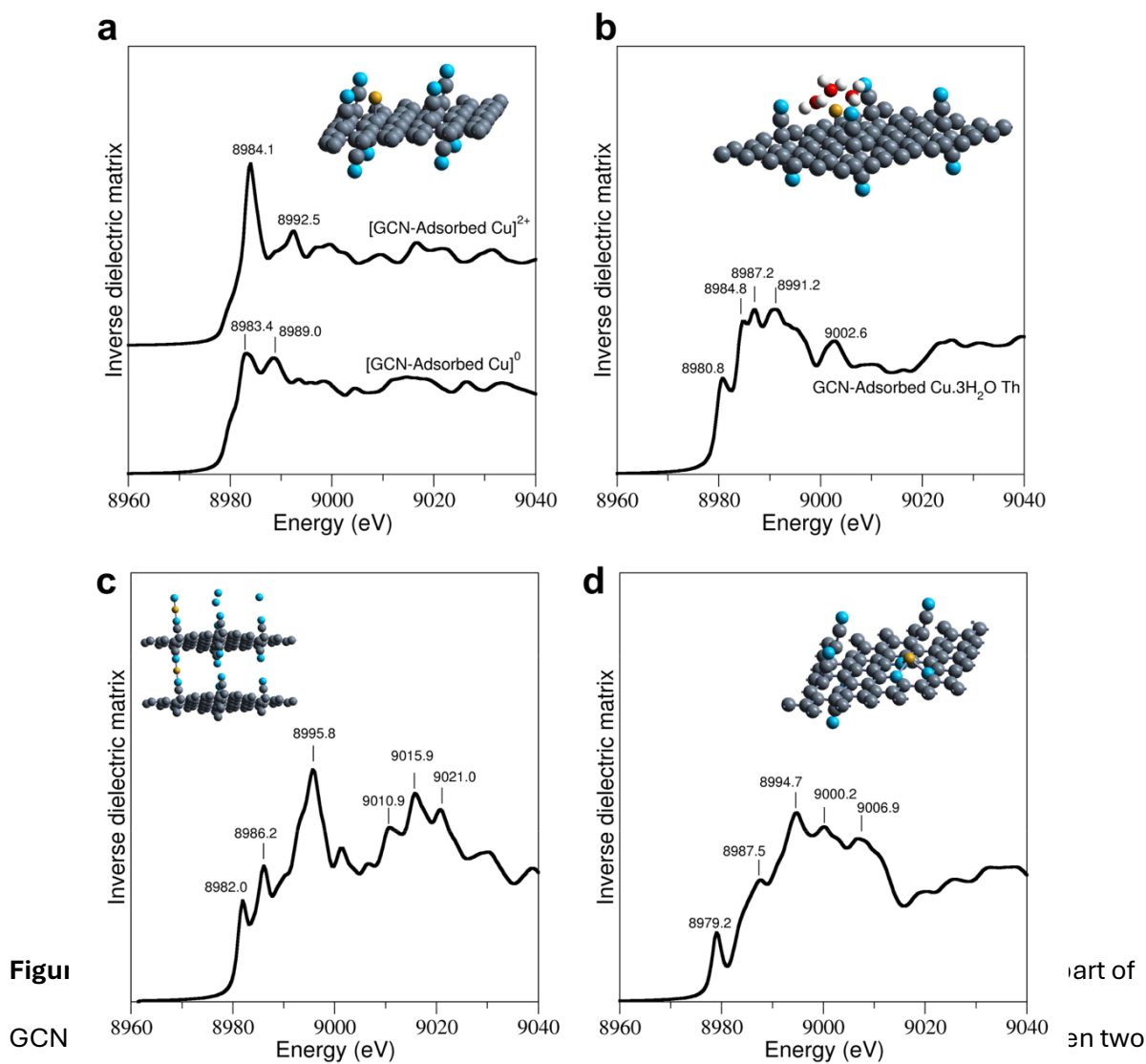

**Figure** [...] part of GCN [...] en two GCN layers, **d**: Cu embedded in a nitrogen-doped defect within GCN. Insets show respective structural models (C atoms in grey, N in blue, Cu in yellow).

Although Cu ions anchored to GCN through nitrile groups represent the most plausible model of GCN-Cu, it cannot be excluded that a fraction of Cu ions may bind to alternative sites, thereby contributing additional features to the overall XANES signal. To account for this possibility, we explore additional configurations: (i) $Cu^{2+}$ ion and $Cu^0$ atom adsorbed on the graphitic carbon backbone of GCN, (ii) Cu atom adsorbed on graphitic part of GCN and coordinated by three water molecules in tetragonal symmetry, (iii) a Cu atom sandwiched between two GCN layers, and (iv) a Cu atom adsorbed at a nitrogen-doped defect in GCN. For the case (i), we use the notation Cu@GCN and $Cu^{2+}$@GCN for zerovalent Cu atom and $Cu^{2+}$ ion adsorbed on graphitic sheet to distinguish them from Cu species anchored through nitrile groups. For Cu@GCN, zerovalent Cu atoms are relatively weakly bonded above one of carbon atoms (interaction energy: -0.596 eV, Cu-C bond length: 2.02 Å), with bonding characterized by a combination of dispersive interaction and dative bonding, similar to other noble metal atoms.[48] As discussed in previous section, only the total charge of the system can be controlled once Cu species are adsorbed. That can be illustrated by the calculated Hirshfeld charges on Cu, which are 0.59 $e$ and 0.37 $e$ for $Cu^{2+}$@GCN and $Cu^0$@GCN, respectively. Nevertheless, the resulting spectra differ; XANES spectrum of $Cu^{2+}$@GCN is dominated by strong peak at 8984.1 eV followed by weaker feature at 8992.5 eV, while the spectrum of $Cu^0$@GCN exhibits weaker features at 8983.4 and 8989.0 eV with similar intensity (Figure 4a). Experimental spectrum of GCN-Cu reported by Bakandritsos and coworkers[15] shows no evidence of signals corresponding to $Cu^{2+}$@GCN and $Cu^0$@GCN, allowing us to rule out surface adsorption as the origin of the Cu(I) oxidation state. Upon coordination with three water molecules in Th symmetry, the spectrum of the Cu atom changes markedly: a new shoulder appears at 8980.8 eV, followed by set of three peaks stacked between 8985 and 8991 eV, and another new peak at 9002.6 eV (Figure 4b). This again highlights the strong influence of the aqueous environment on the electronic structure of Cu species. Indeed, DFT-calculated XANES spectrum has been instrumental in resolving the true structure of $Cu^{2+}$ ions in aqueous systems.[47, 49, 50]

Another possible scenario involves Cu reduction when sandwiched between two GCN layers ($Cu^{2+}$ coordinated by two nitrile group) in multilayer samples. XANES signal of such system is markedly different from that of Cu atom anchored to single nitrile group: its absorption onset shifts to lower energy (8982 eV), followed by peak at 8986.2 eV, and dominant feature at 8995.8 eV. Overall, this spectrum closely resembles that of the linear GCN-Cu.$H_2O$ L1n complex, indicating that the local symmetry of the ligand field around Cu has large effect on the XANES spectrum.

Finally, we examine the spectrum of Cu single atom adsorbed into a nitrogen-doped defect within the GCN lattice. Nitrogen-doped defect consists of a single or double carbon vacancy in the graphitic plane, where the carbon atoms surrounding the vacancy are substituted by nitrogen atoms. It represents a novel type of support for SACs.[51] This structure could plausibly form during the synthesis of GCN-Cu, and higher Cu-N coordination could stabilize a distinct oxidation state. The computed XANES spectrum displays a characteristic low-energy feature at 8979.2 eV (Figure 4d), unique to this configuration and thus serving as clear spectral fingerprint. This also underscores the importance of evaluating absolute energy positions in theoretical XANES simulations, since such diagnostic features would be lost if a spectrum is arbitrarily scaled to match the experimental K-edge onset.

CONCLUSIONS

In summary, the results presented here provide a robust framework for identifying the structural features of Cu-based SACs through theoretical XANES calculations. We established a consistent protocol for simulating XANES spectra that provides the absolute positions of XANES signal along the energy scale. The energy scale was aligned by comparing the excitation energy and Fermi level of each investigated system to those of an isolated Cu atom and by calibrating the calculated spectrum against experiment for a single reference system. Solid-state $CuSO_4 \cdot 5H_2O$ was chosen as this reference system because its calculated spectrum shows excellent agreement with experimental data,[32] which makes the alignment free from any ambiguities. The protocol was validated on a set of standard reference systems, bulk Cu, $Cu_2O$, and CuO, representing Cu(0), Cu(I), and Cu(II) oxidation states, respectively. The calculated spectra reproduced corresponding experimental XANES features with high accuracy, confirming the reliability of the approach.

We then applied this method to investigate XANES characteristics of copper single atoms anchored on cyanographene. The results revealed that assigning an integer oxidation state to the Cu center based solely on the comparison to standard reference systems is problematic, as the electronic structure of the supported Cu atom is significantly influenced by its local coordination and charge delocalization within the GCN support. Moreover, we demonstrated that solvation plays a decisive role: the coordination of Cu by water molecules leads to pronounced changes in the XANES spectra, indicating that the experimental signal must differ substantially between aqueous and vacuum environments.

Finally, our simulations enabled the establishment of clear structure – spectra relationships for various plausible structural motifs in GCN-Cu, providing insight into how coordination geometry and oxidation state may shape the observed spectral features. Overall, this work delivers a comprehensive assessment of both the capabilities and limitations of XANES spectroscopy for characterizing single-atom copper catalysts, offering valuable guidance for distinguishing atomically dispersed Cu species.

ACKNOWLEDGEMENT

This article has been produced with the financial support of the European Union under the REFRESH - Research Excellence for Region Sustainability and High-tech Industries project number CZ.10.03.01/00/22_003/0000048 via the Operational Programme Just Transition. We acknowledge support from the Ministry of Education, Youth and Sports of the Czech Republic through the e-INFRA CZ (ID:90254). We also acknowledge support from ERDF/ESF project TECHSCALE (No. CZ.02.01.01/00/22_008/0004587).

REFERENCES

(1) Gaur, A.; Shrivastava, B. D.; Joshi, S. K. Copper K-edge XANES of Cu(I) and Cu(II) oxide mixtures. *Journal of Physics: Conference Series* **2009**, *190* (1), 012084. DOI: 10.1088/1742-6596/190/1/012084.

(2) Mitchell, S.; Pérez-Ramírez, J. Single atom catalysis: a decade of stunning progress and the promise for a bright future. *Nature Communications* **2020**, *11* (1), 4302. DOI: 10.1038/s41467-020-18182-5.

(3) Qiao, B.; Wang, A.; Yang, X.; Allard, L. F.; Jiang, Z.; Cui, Y.; Liu, J.; Li, J.; Zhang, T. Single-atom catalysis of CO oxidation using Pt1/FeOx. *Nature Chemistry* **2011**, *3* (8), 634-641. DOI: 10.1038/nchem.1095.

(4) Gawande, M. B.; Fornasiero, P.; Zbořil, R. Carbon-Based Single-Atom Catalysts for Advanced Applications. *ACS Catalysis* **2020**, *10* (3), 2231-2259. DOI: 10.1021/acscatal.9b04217.

(5) Wang, Y.; Su, H.; He, Y.; Li, L.; Zhu, S.; Shen, H.; Xie, P.; Fu, X.; Zhou, G.; Feng, C.; et al. Advanced Electrocatalysts with Single-Metal-Atom Active Sites. *Chemical Reviews* **2020**, *120* (21), 12217-12314. DOI: 10.1021/acs.chemrev.0c00594.

(6) Zhang, H.; Jin, X.; Lee, J.-M.; Wang, X. Tailoring of Active Sites from Single to Dual Atom Sites for Highly Efficient Electrocatalysis. *ACS Nano* **2022**, *16* (11), 17572-17592. DOI: 10.1021/acsnano.2c06827.

(7) Cao, L.; Luo, Q.; Liu, W.; Lin, Y.; Liu, X.; Cao, Y.; Zhang, W.; Wu, Y.; Yang, J.; Yao, T.; et al. Identification of single-atom active sites in carbon-based cobalt catalysts during electrocatalytic hydrogen evolution. *Nature Catalysis* **2019**, *2* (2), 134-141. DOI: 10.1038/s41929-018-0203-5.

(8) Kment, Š.; Bakandritsos, A.; Tantis, I.; Kmentová, H.; Zuo, Y.; Henrotte, O.; Naldoni, A.; Otyepka, M.; Varma, R. S.; Zbořil, R. Single Atom Catalysts Based on Earth-Abundant Metals for Energy-Related Applications. *Chemical Reviews* **2024**, *124* (21), 11767-11847. DOI: 10.1021/acs.chemrev.4c00155.

(9) Urso, M.; Ju, X.; Nittoor-Veedu, R.; Lee, H.; Zaoralová, D.; Otyepka, M.; Pumera, M. Single Atom Engineering for Electrocatalysis: Fundamentals and Applications. *ACS Catalysis* **2025**, *15* (13), 11617-11663. DOI: 10.1021/acscatal.4c08027.

(10) Maurer, F.; Jelic, J.; Wang, J.; Gänzler, A.; Dolcet, P.; Wöll, C.; Wang, Y.; Studt, F.; Casapu, M.; Grunwaldt, J.-D. Tracking the formation, fate and consequence for catalytic activity of Pt single sites on CeO2. *Nature Catalysis* **2020**, *3* (10), 824-833. DOI: 10.1038/s41929-020-00508-7.

(11) Sarma, B. B.; Maurer, F.; Doronkin, D. E.; Grunwaldt, J.-D. Design of Single-Atom Catalysts and Tracking Their Fate Using Operando and Advanced X-ray Spectroscopic Tools. *Chemical Reviews* **2023**, *123* (1), 379-444. DOI: 10.1021/acs.chemrev.2c00495.

(12) Nitopi, S.; Bertheussen, E.; Scott, S. B.; Liu, X.; Engstfeld, A. K.; Horch, S.; Seger, B.; Stephens, I. E. L.; Chan, K.; Hahn, C.; et al. Progress and Perspectives of Electrochemical CO2 Reduction on Copper in Aqueous Electrolyte. *Chemical Reviews* **2019**, *119* (12), 7610-7672. DOI: 10.1021/acs.chemrev.8b00705.

(13) Chang, F.; Xiao, M.; Miao, R.; Liu, Y.; Ren, M.; Jia, Z.; Han, D.; Yuan, Y.; Bai, Z.; Yang, L. Copper-Based Catalysts for Electrochemical Carbon Dioxide Reduction to Multicarbon Products. *Electrochemical Energy Reviews* **2022**, *5* (3), 4. DOI: 10.1007/s41918-022-00139-5.

(14) Zhang, Y.; Zhao, J.; Wang, H.; Xiao, B.; Zhang, W.; Zhao, X.; Lv, T.; Thangamuthu, M.; Zhang, J.; Guo, Y.; et al. Single-atom Cu anchored catalysts for photocatalytic renewable H2 production with a quantum efficiency of 56%. *Nature Communications* **2022**, *13* (1), 58. DOI: 10.1038/s41467-021-27698-3.

(15) Bakandritsos, A.; Kadam, R. G.; Kumar, P.; Zoppellaro, G.; Medved, M.; Tuček, J.; Montini, T.; Tomanec, O.; Andrýsková, P.; Drahoš, B.; et al. Mixed-Valence Single-Atom Catalyst Derived

from Functionalized Graphene. *Advanced Materials* **2019**, *31* (17), 1900323. DOI: 10.1002/adma.201900323.

(16) Gazis, T. A.; Palit, S.; Cipriano, L. A.; Allasia, N.; Collins, S. M.; Ramasse, Q. M.; Kwon, I. S.; Sterrer, M.; Di Liberto, G.; Vilé, G. Copper Single-Atom Catalyst for Efficient C–S Coupling in Thioether Synthesis. *Angewandte Chemie International Edition* **2025**, *64* (38), e202510632. DOI: 10.1002/anie.202510632.

(17) Lazar, P.; Mach, R.; Otyepka, M. Spectroscopic Fingerprints of Graphitic, Pyrrolic, Pyridinic, and Chemisorbed Nitrogen in N-Doped Graphene. *The Journal of Physical Chemistry C* **2019**, *123* (16), 10695-10702. DOI: 10.1021/acs.jpcc.9b02163.

(18) Lazar, P.; Hrubý, V.; Petr, M.; Baďura, Z.; Zoppellaro, G.; Otyepka, M. Decoding structural characteristics of fluorinated graphene via Computer-Aided spectroscopic analysis. *Carbon* **2025**, *243*, 120567. DOI: 10.1016/j.carbon.2025.120567.

(19) Susi, T.; Kaukonen, M.; Havu, P.; Ljungberg, M. P.; Ayala, P.; Kauppinen, E. I. Core level binding energies of functionalized and defective graphene. *Beilstein Journal of Nanotechnology* **2014**, *5*, 121-132. DOI: 10.3762/bjnano.5.12.

(20) Salcedo, A.; Zengel, D.; Maurer, F.; Casapu, M.; Grunwaldt, J.-D.; Michel, C.; Loffreda, D. Identifying the Structure of Supported Metal Catalysts Using Vibrational Fingerprints from Ab Initio Nanoscale Models. *Small* **2023**, *19* (34), 2300945. DOI: 10.1002/smll.202300945 (accessed 2025/06/24).

(21) Thang, H. V.; Pacchioni, G.; DeRita, L.; Christopher, P. Nature of stable single atom Pt catalysts dispersed on anatase TiO2. *Journal of Catalysis* **2018**, *367*, 104-114. DOI: 10.1016/j.jcat.2018.08.025.

(22) DeRita, L.; Resasco, J.; Dai, S.; Boubnov, A.; Thang, H. V.; Hoffman, A. S.; Ro, I.; Graham, G. W.; Bare, S. R.; Pacchioni, G.; et al. Structural evolution of atomically dispersed Pt catalysts dictates reactivity. *Nature Materials* **2019**, *18* (7), 746-751. DOI: 10.1038/s41563-019-0349-9.

(23) Kraushofer, F.; Parkinson, G. S. Single-Atom Catalysis: Insights from Model Systems. *Chemical Reviews* **2022**, *122* (18), 14911-14939. DOI: 10.1021/acs.chemrev.2c00259.

(24) Blöchl, P. E. Projector augmented-wave method. *Physical Review B* **1994**, *50* (24), 17953-17979. DOI: 10.1103/PhysRevB.50.17953.

(25) Kresse, G.; Joubert, D. From ultrasoft pseudopotentials to the projector augmented-wave method. *Physical Review B* **1999**, *59* (3), 1758-1775. DOI: 10.1103/PhysRevB.59.1758.

(26) Klimeš, J.; Bowler, D. R.; Michaelides, A. Van der Waals density functionals applied to solids. *Physical Review B* **2011**, *83* (19), 195313. DOI: 10.1103/PhysRevB.83.195131.

(27) Rehr, J. J.; Kas, J. J.; Vila, F. D.; Prange, M. P.; Jorissen, K. Parameter-free calculations of X-ray spectra with FEFF9. *Physical Chemistry Chemical Physics* **2010**, *12* (21), 5503-5513, 10.1039/B926434E. DOI: 10.1039/B926434E.

(28) Bunău, O.; Joly, Y. Self-consistent aspects of x-ray absorption calculations. *Journal of Physics: Condensed Matter* **2009**, *21* (34), 345501. DOI: 10.1088/0953-8984/21/34/345501.

(29) Chen, Y.; Chen, C.; Hwang, I.; Davis, M. J.; Yang, W.; Sun, C.; Lee, G.-H.; McReynolds, D.; Allan, D.; Marulanda Arias, J.; et al. Robust Machine Learning Inference from X-ray Absorption Near Edge Spectra through Featurization. *Chemistry of Materials* **2024**, *36* (5), 2304-2313. DOI: 10.1021/acs.chemmater.3c02584.

(30) Guo, H.; Carbone, M. R.; Cao, C.; Qu, J.; Du, Y.; Bak, S.-M.; Weiland, C.; Wang, F.; Yoo, S.; Artrith, N.; et al. Simulated sulfur K-edge X-ray absorption spectroscopy database of lithium thiophosphate solid electrolytes. *Scientific Data* **2023**, *10* (1), 349. DOI: 10.1038/s41597-023-02262-4.

(31) England, A. H.; Duffin, A. M.; Schwartz, C. P.; Uejio, J. S.; Prendergast, D.; Saykally, R. J. On the hydration and hydrolysis of carbon dioxide. *Chemical Physics Letters* **2011**, *514* (4), 187-195. DOI: 10.1016/j.cplett.2011.08.063.

(32) Mongioví, C.; Crini, G.; Gabrion, X.; Placet, V.; Blondeau-Patissier, V.; Krystianiak, A.; Durand, S.; Beaugrand, J.; Dorlando, A.; Rivard, C.; et al. Revealing the adsorption mechanism of copper on hemp-based materials through EDX, nano-CT, XPS, FTIR, Raman, and XANES characterization techniques. *Chemical Engineering Journal Advances* **2022**, *10*, 100282. DOI: 10.1016/j.ceja.2022.100282.

(33) Wang, Y.; Lany, S.; Ghanbaja, J.; Fagot-Revurat, Y.; Chen, Y. P.; Soldera, F.; Horwat, D.; Mücklich, F.; Pierson, J. F. Electronic structures of Cu2O,Cu4O3, and CuO: A joint experimental and theoretical study. *Physical Review B* **2016**, *94* (24), 245418. DOI: 10.1103/PhysRevB.94.245418.

(34) Yang, B. X.; Thurston, T. R.; Tranquada, J. M.; Shirane, G. Magnetic neutron scattering study of single-crystal cupric oxide. *Physical Review B* **1989**, *39* (7), 4343-4349. DOI: 10.1103/PhysRevB.39.4343.

(35) Forsyth, J. B.; Brown, P. J.; Wanklyn, B. M. Magnetism in cupric oxide. *Journal of Physics C: Solid State Physics* **1988**, *21* (15), 2917. DOI: 10.1088/0022-3719/21/15/023.

(36) Živković, A.; Roldan, A.; de Leeuw, N. H. Density functional theory study explaining the underperformance of copper oxides as photovoltaic absorbers. *Physical Review B* **2019**, *99* (3), 035154. DOI: 10.1103/PhysRevB.99.035154.

(37) Wu, D.; Zhang, Q.; Tao, M. LSDA+U study of cupric oxide: Electronic structure and native point defects. *Physical Review B* **2006**, *73* (23), 235206. DOI: 10.1103/PhysRevB.73.235206.

(38) Heinemann, M.; Eifert, B.; Heiliger, C. Band structure and phase stability of the copper oxides Cu2O, CuO, and Cu4O3. *Physical Review B* **2013**, *87* (11), 115111. DOI: 10.1103/PhysRevB.87.115111.

(39) Rödl, C.; Sottile, F.; Reining, L. Quasiparticle excitations in the photoemission spectrum of CuO from first principles: A $GW$ study. *Physical Review B* **2015**, *91* (4), 045102. DOI: 10.1103/PhysRevB.91.045102.

(40) Asbrink, S.; Norrby, L.-J. A refinement of the crystal structure of copper(II) oxide with a discussion of some exceptional e.s.d.'s. *Acta Crystallographica Section B* **1970**, *26* (1), 8-15. DOI: doi:10.1107/S0567740870001838.

(41) Cuartero, V.; Monteseguro, V.; Otero-de-la-Roza, A.; El Idrissi, M.; Mathon, O.; Shinmei, T.; Irifune, T.; Sanson, A. Interplay between local structure, vibrational and electronic properties on CuO under pressure. *Physical Chemistry Chemical Physics* **2020**, *22* (42), 24299-24309, 10.1039/D0CP04878J. DOI: 10.1039/D0CP04878J.

(42) Sinha, R. N.; Mahto, P.; Chetal, A. R. X-rayK-absorption edge interpretation and energy determination. *Zeitschrift für Physik B Condensed Matter* **1990**, *81* (2), 229-232. DOI: 10.1007/BF01309353.

(43) Bocharov, S.; Kirchner, T.; Dräger, G.; Šipr, O.; Šimůnek, A. Dipole and quadrupole contributions to polarized Cu K x-ray absorption near-edge structure spectra of CuO. *Physical Review B* **2001**, *63* (4), 045104. DOI: 10.1103/PhysRevB.63.045104.

(44) Zaoralová, D.; Mach, R.; Lazar, P.; Medveď, M.; Otyepka, M. Anchoring of Transition Metals to Graphene Derivatives as an Efficient Approach for Designing Single-Atom Catalysts. *Advanced Materials Interfaces* **2021**, *8* (8), 2001392. DOI: 10.1002/admi.202001392 (accessed 2025/06/16).

(45) Průcha, R.; Hrubý, V.; Zaoralová, D.; Otyepková, E.; Šedajová, V.; Kolařík, J.; Zbořil, R.; Medved', M.; Otyepka, M. Coordination effects on the binding of late 3d single metal species to cyanographene. *Physical Chemistry Chemical Physics* **2023**, *25* (1), 286-296. DOI: 10.1039/D2CP04076J.

(46) Kuzmin, A.; Timoshenko, J.; Kalinko, A.; Jonane, I.; Anspoks, A. Treatment of disorder effects in X-ray absorption spectra beyond the conventional approach. *Radiation Physics and Chemistry* **2020**, *175*, 108112. DOI: 10.1016/j.radphyschem.2018.12.032.

(47) La Penna, G.; Minicozzi, V.; Morante, S.; Rossi, G. C.; Stellato, F. A first-principle calculation of the XANES spectrum of Cu2+ in water. *The Journal of Chemical Physics* **2015**, *143* (12), 124508. DOI: 10.1063/1.4931808.

(48) Granatier, J.; Lazar, P.; Otyepka, M.; Hobza, P. The Nature of the Binding of Au, Ag, and Pd to Benzene, Coronene, and Graphene: From Benchmark CCSD(T) Calculations to Plane-Wave

DFT Calculations. *Journal of Chemical Theory and Computation* **2011**, *7* (11), 3743-3755. DOI: 10.1021/ct200625h.

(49) Frank, P.; Benfatto, M.; Szilagyi, R. K.; D'Angelo, P.; Della Longa, S.; Hodgson, K. O. The Solution Structure of [Cu(aq)]2+ and Its Implications for Rack-Induced Bonding in Blue Copper Protein Active Sites. *Inorganic Chemistry* **2005**, *44* (6), 1922-1933. DOI: 10.1021/ic0400639.

(50) Frank, P.; Benfatto, M.; Qayyum, M. [Cu(aq)]2+ is structurally plastic and the axially elongated octahedron goes missing. *The Journal of Chemical Physics* **2018**, *148* (20), 204302. DOI: 10.1063/1.5024693 (accessed 8/31/2025).

(51) Zaoralová, D.; Langer, R.; Otyepka, M. Iron Single-Atom Catalysts Anchored on Defect-Engineered N-Doped Graphene Reveal an Interplay between CO2 Reduction Activity and Stability. *ACS Sustainable Chemistry & Engineering* **2025**, *13* (22), 8319-8330. DOI: 10.1021/acssuschemeng.5c01417.